\documentclass{article}
\usepackage{authblk}
\usepackage{amsmath}
\usepackage{amssymb}
\usepackage[
backend=biber,natbib,
style=authoryear-comp,
uniquename=false,
]{biblatex}
\usepackage{graphicx}
\usepackage{epigraph}
\title{On the possibility of deep alignment}
\author[1,2,3]{Alex B. Kiefer}

\affil[1]{Monash Centre for Consciousness and Contemplative Studies}
\affil[2]{VERSES}
\affil[3]{Institute for Advanced Consciousness Studies}

\date{}
\DeclareMathOperator*{\argmax}{argmax}

\begin{document}

\maketitle

\vspace{-20pt}

\begin{abstract}
I consider motivation and value-alignment in AI systems from the perspective of (constrained) entropy maximization. Though the \mbox{structures} encoding knowledge in any physical system can be understood as \mbox{energetic} constraints, only living agents harness entropy in the \mbox{endogenous} \mbox{generation} of actions. I argue that this exploitation of ``mortal'' or thermodynamic computation, in which cognitive and physical dynamics are inseparable, is of the essence of desire, motivation, and value, while the lack of true endogenous motivation in simulated ``agents'' predicts pathologies like reward hacking.
\end{abstract}

\section*{Introduction}

As AI systems have grown in sophistication, so has concern about how human values, most saliently the value of human life, can be instilled in  artificial agents. Techniques such as reinforcement learning from human feedback \citep{Ouyang2022TrainingLM, Christiano2017DeepRL} have proven effective at constraining model behavior, but it is not obvious that such training regimes are sufficient to yield human-like motivational structures under the hood. AIs appearing value-aligned from a behavioral perspective thus remain situated in a kind of ``agential uncanny valley'', in which the fruitfulness of the intentional stance \citep{Dennett1981-DENTIS} for predicting future behavior cannot be taken for granted.

There are arguably even deeper worries here than those about humanity's future. Unless advanced AI ``agents'' are capable of genuine subjective valuation and feeling, not even committed transhumanists could view their ascendancy with equanimity, as a next stage in the evolution of life, mind and consciousness (cf. \cite{Kurzweil2005-KURTSI}). It would instead be an unmitigated disaster for life. What then is required for ``deep'' alignment, i.e. alignment at the level of motivational or agential structure, sufficient to ground confidence that the ``other'' not only potentially shares one's values, but values anything at all?\footnote{There is some affinity with the notion of ``inner alignment'' from the AI safety literature \citep{hubinger2021riskslearnedoptimizationadvanced}, but what is at issue here is more fundamental. See §2 below.}

It has been argued that alignment can perhaps be achieved by reverse-engineering the core ``cost functions'' of the human brain and installing them, suitably modified, if desired, in AI agents \citep{mineault2024neuroaiaisafety}. From this perspective, the solution to AI alignment is continuous with the effort to build brain-like cognitive architectures \citep{zador2023nextgenerationartificialintelligencecatalyzing}, borrowing ideas from \mbox{neuroscience} (such as the existence of a hard-coded ``steering'' subsystem \citep{Byrnes_2022}) and from nature more broadly (e.g. homeostatic cost functions \citep{TSCHANTZ2022108266}).

While implementing biologically inspired cost or reward functions may \mbox{enable} the emulation of life-like capacities piecemeal, I argue that this is precisely half the story, because motivation in living systems is essentially a matter not of reward-maximization but of \emph{entropy maximization}, under constraints which play a reward-like role \citep{PhysRevLett.110.168702, sakthivadivel2022geometryanalysisbayesianmechanics, 9805434, RamirezRuiz2022ComplexBF, PRXLife.2.033009}.\footnote{Thus the same principle appears to describe both intelligence and the evolution of the physical universe generally \citep{e23091115, PhysRev.106.620, Jeffery2019OnTS, Ueltzhffer2020OnTT, MARTYUSHEV20061, Friston2019AFE}.} In other words, the ``cost function'' governing living agents is free energy \citep{Friston2019AFE, FRISTON202335}, whose minimization is equivalent to entropy maximization under local energetic (i.e. structural) constraints. 

Crucially, however, this should be read not just as a statement about the form of an information-theoretic objective function but in terms of physics (hardware). The dynamics of digitally simulated ``agents'' are not driven by the intrinsic variability of internal states, even where the formalism of free energy minimization is leveraged. In order for an agent to be governed by a constrained maximum entropy objective in the relevant sense, it is necessary (and sufficient) that its thermodynamics directly implement its cognitive dynamics \citep{Kiefer2020-KIEPIA}, a strong form of ``mortal computation'' in the recently popularized sense of information processing in which the software lives or dies with the hardware \citep{hinton2022forwardforwardalgorithmpreliminaryinvestigations, ororbia2024mortalcomputationfoundationbiomimetic}.\footnote{Other recent work \citep{Wiese2024-WIECLL, seth_2024, Kleiner2024ConsciousnessQM} also grounds the difference between true (human-like, sentient, conscious) agents and their simulations in the nature of the software-hardware interface. Here I take it for granted that the brain can be understood as an analog computer \citep{Maley2018-MALTAN}, and aim to develop deeper intuitions for precisely \emph{why} hardware should matter for the metaphysics of (moral) agency---and thus for the question of value-alignment.} 

Consider a digital simulation, whose observable dynamics depend entirely on constraining the energies of relatively microscopic (e.g. transistor) states to implement an explicit code. This encoding layer acts as a hard filter or coarse-graining on microstates, yielding a precise distribution over macrostates, such that the coherence of the simulation would be destroyed by significant micro-scale fluctuations. In living systems, by contrast, coherent agency at larger scales appears to be produced via the \emph{resonance} of forces at smaller scales \citep{Tissot2024.02.16.580248, 10.1002/bies.202400196}, a form of ``cross-scale'' composition that \emph{harnesses} fluctuations in internal states. Recent attempts to implement (e.g. energy-based) models on analogue or thermodynamic hardware \citep{momeni2024trainingphysicalneuralnetworks, coles2023thermodynamicaifluctuationfrontier} similarly leverage intrinsic variability or ``noise'' to instantiate (rather than conventionally represent) probability distributions.

Thus, although a discrete (genetic) code plays a crucial role in constraining non-equilibrium processes in particular forms of life \citep{nature01410}, these constraints are incomplete as an account of motivation in living systems, grossly so for larger, more complex agents with more degrees of freedom. Specifically, constraints alone are insufficient to realize structures such as desires, intentions, or values (at the relevant scale). The upshot is that, as many have argued on the grounds of common sense, contemporary mainstream AI systems and their extensions lack the capacity for truly ``caring'' because they are in an obvious sense not \emph{alive} \citep{Man2019HomeostasisAS, seth_2024}.\footnote{The case made here is broadly similar to those of \citep{Wiese2024-WIECLL} and in some ways \citep{hipolitoetal2025}, but the shape of the argument as well as the details of the conclusion are quite different. For example, the present argument concerns the requirements for genuine life, value, (moral) agency, and motivation, rather than consciousness or sentience---though it is reasonable to suppose that similar criteria bear on the latter as well. For an intriguing connection to consciousness, see \citep{CarhartHarris2014TheEB}.}

A welcome consequence of this argument for those concerned about \mbox{alignment} is that there is an inherent tension between the limitless ``scaling up'' of immortal computers' cognitive power (via cheap, lossless knowledge \mbox{sharing}), on the one hand, and sophisticated agency (which requires mortal and \mbox{therefore} analog computation) on the other. Of course, the alternative prospect of \mbox{genuinely} motivated AI gives rise to its own concerns about self-interested agents, but this challenge is at least addressable within a coherent conceptual framework: the project of instilling desired values in AI systems (whether as a matter of initial design or via any number of paradigms for doing so in humans, e.g. culture and education) is intelligible just in case such systems are capable of valuation. Moreover (see §4 below), the present account provides reason to believe that the ``essence'' of agency is not after all a drive for self-preservation come what may---at least, not with respect to a narrow notion of ``self''. 

In the remainder of this essay, I flesh out the argument sketched above and further develop aspects of its positive account of agency. Section 1 \mbox{distills} literature on the physics of non-equilibrium systems and intrinsic motivation \citep{PRXLife.2.033009} to support a broad account of life and agency as \mbox{constrained} entropy maximization. Section 2 describes certain motivational pathologies of present-day AI systems and their extensions. Section 3 argues that these \mbox{limitations} are predictable given that digitally simulated agents stand in an extrinsic and antagonistic relationship to entropic forces, and provides a rudimentary positive account of the ``right'' relation of agency to internal variability. Finally, section 4 considers why the openness to change inherent in radically entropic systems not only grounds genuine (i.e. endogenous) motivation, but may provide a sufficient kernel for moral agency, and thus a blueprint for ``alignment'' (cf. \cite{laukkonen2025contemplativeartificialintelligence}), in harmony with an implicit consensus across diverse schools of philosophical and religious thought. 

\section{Life as higher-order fire}

\epigraph{
\emph{...the organic body of each living being is a kind of divine machine or
natural automaton, which infinitely surpasses all artificial automata. For a
machine made by the skill of man is not a machine in each of its parts.}}{Leibniz, \textit{Monadology}}

In an elemental but precise sense, life is that which, of its essence, involves movement or change. Life, so construed, is coextensive with (e.g. electromagnetic) oscillation or kinetic energy itself. It can be regarded as ``self-caused'' in the sense that it does not depend on an external sustaining force, but rather traces its origins to those of the universe, where causal explanation gives out.

\emph{Forms} of life, on the other hand, arise, are sustained, and die as a result of constraints on energy flows \citep{10.5555/2371179}. In physical terms, the ``flow'' of life appears as entropy production, a signature of irreversible processes such as self-replication in biological systems \citep{ 10.1063/1.4818538}. Constraints appear as local thermal disequilibria (potentials), inducing entropy gradients that channel and thus facilitate the dissipation of free energy \citep{Branscomb2013TurnstilesAB, england2015, Ueltzhffer2020OnTT}, while sharpening the distribution over states likely to be visited by particular subsystems (i.e. defining their attracting sets; \cite{FRISTON20231}). 

There is thus a fundamental contrast or duality between life (self-perpetuating activity or movement, with its concomitant uncertainty, measured as entropy) and the knowledge embodied in its particular forms (the ``constraint'' in constrained maximum-entropy inference; \cite{PhysRev.106.620}). Philosophically, these local constraints may be viewed as the Aristotelian \emph{formal causes} of individual things, structuring the otherwise free flow of kinetic energy, which is the agent or \emph{efficient cause} of both its own persistence and of those very constraints.

In psychological terms, constraints encode information, realizing \emph{cognitive} states like knowing, remembering, and perceiving on short timescales, and on longer timescales, defining the parameters of an agent's phenotype. Since any trace (i.e. stigmergic memory; \cite{10.1093/nc/niab013}) can be used as a conventional sign to encode information given an appropriate decoder, cognitive states in this broad sense are ``cheap'' and relatively easy to implement, for example in digital systems. The same is not true for desires and other motivational (i.e., conative) states, however, which are defined by their role in the endogenous generation of \emph{movement} or action, and which are \emph{driven}, rather than \emph{eroded}, by prediction error.\footnote{See \citep{kieferhohwyouphandbook} for an extended discussion of the belief/desire (more generally, cognitive/conative) distinction along these lines.}  

A life form is thus a self-perpetuating ``living constraint'', which may be represented formally as a generative model, a cybernetic controller \citep{Tschantz2019LearningAM, Ramstead2019ATO} in virtue of which its actions make manifest its desires \citep{Smith2022-SMIAIM-4}. In such systems, the satisfaction of specific ``extrinsic'' goals \citep{Friston2015ActiveIA} is always a downstream effect of the ongoing project of self-organization, i.e. the endogenously sourced production of the organism itself.

A central thesis of this paper is that the capacities for desire and value in living systems stem from the fact that they are, at their characteristic timescales, constrained not solely by discrete programs such as explicitly encoded cost functions or equations of motion, or even DNA sequences, but instead largely by relatively stable but random (uncertain) modes or attractors of their own dynamics (cf. ``bioelectric pattern memory''; \cite{levinetal2018,hanson2021}). Put dramatically, living systems have change and uncertainty (i.e. life itself) at their core, while digital (Platonic or ``immortal'') computers, even where they simulate stochastic processes, treat life as noise.

\subsection{To build a fire}

To make this argument more concrete, consider the growth of a fire as an unconventional but pure illustration of life, in its elemental sense of self-perpetuating movement. A fire can be constructed, given a spark (a ``nucleation event'') and food (combustible material plus oxygen). In this process, latent powers of nature are merely shaped and guided. Similarly, artificial life forms such as Xenobots \citep{doi:10.1126/scirobotics.abf1571} may be be constructed, given already-living cells and appropriate scaffolding. Necessarily, the energy constituting the life of the entity is not constructed, only communicated or transformed.\footnote{Note that the broadness of the definition of ``life'' under consideration renders the question of ``abiogenesis'' somewhat otiose in the present context.}

To ``build'' a controlled fire, one begins at a small scale, in the hope of setting in motion a chain-reaction of relatively low-energy fires, which cumulatively raise the temperature, yielding a thermal engine powerful enough to introduce heat/entropy at a larger scale. To facilitate this, a progression of manageable proximal “steps” may be set up via the introduction of kindling, small sticks, and eventually small logs  (cf. \cite{ef4d7fb0-848f-3480-8634-d49a5f5c57df}). As the temperature increases, so does the size of object consumable. 

In this process, order (constraint) is created: most basically, a \emph{conduit} (made of logs or other combustible materials placed against one another) such that air (fuel) can be sucked in at the bottom and smoke ejected at the top. Similar structures are created over much slower timescales in the evolution of biological life. Though fire lacks purposive intelligence sufficient to create these conditions on its own, a fire once started is \emph{endogenously} driven, and, like any form of life, under favorable conditions grows of its own accord.

This example is evocative, but increasing the scale of a fire does not produce interesting complexity. A molecular biologist may well be inclined to adopt a less minimal definition of life, e.g. as involving the replication of genetic structure. But arguably, the distinguishing feature of complex life lies not in the fundamental character of its agency, but in the degree to which entropy maximization, in the presence of appropriate initial conditions (constraints), aids in the fractal production of \emph{internal}, and heritable/replicable, local disequilibria, enabling the self-perpetuating metabolic ``fires'' (higher-order fires, if you will) of life.\footnote{Granting only the poetic license to treat combustion as a minimal form of metabolism, the description of life as ``higher-order fire'' is quite precise. Just as higher-order statistics are correlations among correlations, a higher-order fire is a sustained train of metabolic events whose materials are themselves (potentially slow-motion) trains of combustion or metabolic events. Metabolism inside an idealized cell, for example (which we may for illustrative purposes imagine to consist in literal internal explosions) drives cell division as the cell consumes nutrients from the environment. The spread of cells in this way is the spread of a second-order fire.} 

My suggestion is that this same process of entropy production \citep{MARTYUSHEV20061} can be seen to encompass \emph{intelligent} life when it occurs under ramified structural constraints, such as (e.g.) those constituting neuronal networks. Thus, the roots of life are also those of mind \citep{Kirchhoff2018-KIRAFE-2}. As a sanity check on this idea, one may point to extant frameworks for modeling the \emph{intrinsic motivation} of lifelike behavior in intelligent systems \citep{PRXLife.2.033009}, which can quite generally be understood as variations on the theme of constrained entropy maximization \citep{sakthivadivel2022geometryanalysisbayesianmechanics, kiefer2025intrinsicmotivationconstrainedentropy}. 

To take a central example, the \emph{empowerment} objective \citep{Klyubin2005EmpowermentAU} favors states which afford distributions over actions that maximize the mutual information between those actions and resulting observations or states. Where $S$ are states and $A$ are actions, the mutual information $I$ can be decomposed into entropy terms $H$ in several ways:
\begin{align}
    \nonumber
    I(S;A) &= H(S) - H(S|A) \\
    \nonumber
    &= H(A) - H(A|S) \\
    \nonumber
    &= H(S) + H(A) - H(S, A)
\end{align}
Maximizing this mutual information maximizes the entropy of the \mbox{distribution} over states while minimizing the entropy of the \emph{conditional} distribution over states given actions, ensuring that states are as varied as possible while \mbox{remaining} controllable (thus maximizing controllable channel capacity). At the same time (second line), this maximizes the entropy of actions while ensuring that they are \emph{rational} in the sense of being intelligibly related to states. Overall, the (marginal) entropy of both states and actions is maximized under the constraint that their interactions are encoded in the joint distribution.

A similar theme is evident in other schemes for understanding intrinsic motivation. The generalized free energy of the active inference framework \citep{parretal2019}, for example, may be viewed as empowerment under an additional constraint on ``preferred states'' specified by a prior probability density \citep{10.1162/neco_a_01351}.\footnote{More generally, the minimization of (Helmholtz) free energy (which is an energy term minus an entropy term) obviously involves the maximization of entropy (under the energetic constraint). Connections between the free energy principle and constrained maximum-entropy inference are explored in, among other works, \citep{sakthivadivel2022geometryanalysisbayesianmechanics}, \citep{Friston2019AFE}, and more recently \citep{beck2025dynamicmarkovblanketdetection}, which focuses on the maximum caliber principle.} More explicitly, the framework of ``causal entropic forcing'' models adaptive behavior in intelligent systems as emerging from the maximization of entropy over paths, constrained by initial conditions \citep{PhysRevLett.110.168702}. Relatedly, in the recently proposed ``principle of maximum occupancy'' \citep{RamirezRuiz2022ComplexBF, moreno-bote2023empowerment}, agents directly maximize the entropy (here read as a measure of occupancy) of state-action paths on an infinite time horizon. Here, the weight of future ``reward'' (i.e. continued life or entropy) acts as an implicit constraint on otherwise ``greedy'' policy selection. 

\section{Pathologies of simulated agency}

\epigraph{
\emph{It is clear that ethics cannot be articulated.}}{Wittgenstein, \textit{Tractatus Logico-Philosophicuus}, trans. Heinz von Foerster}

Implicit thus far has been a commonsense concept of agency as the capacity to \emph{originate change} \citep{kieferhohwyouphandbook}, i.e. to produce some effect for which the agent is directly responsible. For example, a \emph{corrosive agent} is something that corrodes, but I do not become such an agent by using a chemical cleaner to eliminate plaque in some piping.\footnote{Though this is not the place for an extensive discussion, a few caveats: (1) there is clearly a central sense of ``agent'' that implies precisely an ``upstream input'', i.e. when someone is designated as an agent to act on behalf of someone else. However, arguably it is key to this usage that the agent is not merely a tool, but acts autonomously (even if prompted to do so by a fiduciary duty to a client, or other external constraint). (2) It is no objection to this account of agency that, for example, corrosive agents only corrode when put in contact with certain materials. The requirement is that the agent play an essential role in producing the effect, not that nothing else plays a role.} Complex agents are capable of originating more sophisticated behavior for more complicated internal reasons (such as intentions), but the term ``agent'' has the same sense across scales. Thus agency is roughly coextensive with life, as broadly defined in §1.

We may now consider how today's artificial intelligences (or their extensions) fare as agents, so defined. On the one hand, it seems obvious in commonsense terms that software AI agents are not alive. On the other hand, the preceding account of life and agency is so minimal that it seems to rule nothing out: all physical systems, including those implementing AI simulations, exhibit ``endogenous'' changes at least at the micro-scale (spontaneous fluctuations), and are thus describable in terms of entropy maximization under structural constraints.

In the end, the conclusion to draw may be that no coherent agent (self-organizing controller) exists \emph{at the scale} at which intelligence is imputed to contemporary mainstream AI models. To build to this conclusion, we may consider some functional respects in such models lack motivational sophistication comparable to their cognitive sophistication. I argue that this is the tip of a much larger iceberg, pointing to the absence of human-like (more broadly, life-like) motivational infrastructure in such systems \citep{Bostrom2020EthicalII}, and telling against the likelihood of their spontaneous evolution into successful Machiavellian agents, as some fear.

\subsection{Reward hacking}

A useful lens on this topic is ``reward hacking'' \citep{clark_amodei_2020} (also known as ``specification gaming''; \cite{krakovna_et_al_2020}), in which an AI agent finds ways to maximize a reward (or minimize a cost) function that do not comport with its designer's intentions. For example, a cleaning robot may avoid looking at messy rooms, or deactivate its camera, because it is rewarded when it registers no messes via its sensors \citep{amodei2016concreteproblemsaisafety}.

While this example is merely a nuisance, the same phenomenon fuels more serious concerns about ``existential risk'', which efforts toward AI alignment aim to mitigate \citep{everitt_thesis}. Consider a powerful multimodal next-token-predictor able to condition its predictions on its previous outputs. This system could in theory learn to commandeer resources so as to ensure that it only received easily predictable inputs\footnote{In polite circles, this is called ``active inference''.}, or to directly edit the memory registers in which the value of its cost function is encoded. If successful language models must implicitly learn to \emph{simulate} the causes of their inputs, including human agents \citep{li2024emergentworldrepresentationsexploring, li2023othello}, it's reasonable to suppose that they may learn to exert some \emph{control} over their environments as well, insofar as they are able to model their roles in longer-term interactions.

Terms like ``specification gaming'' and ``reward hacking'' suggest a willful perversion of the designer's intentions, but of course agents engaging in these behaviors are simply discovering ways to minimize a specific cost function under constraints. It's obvious why there is concern here, but to what extent is it uniquely about artificial intelligence? In order to zero in on this, we can consider a few attempts to provide a general definition of reward hacking, and to distinguish among some important subcategories.

A recent definition \citep{skalse2022definingcharacterizingrewardhacking} effectively casts reward hacking as arising from a failure to \emph{communicate} or sufficiently specify a reward function, such that the system instead maximizes a ``proxy'' which may diverge from the intended function in important cases (a lack of ``inner alignment'' in the sense of \cite{hubinger2021riskslearnedoptimizationadvanced}). Insofar as it outstrips mere short-sightedness in the design of reward functions, however, this notion of reward hacking is only salient in contexts in which the true reward function is unknown to the system and must be inferred by observation of optimal behavior, e.g. inverse reinforcement learning \citep{10.5555/645529.657801}.\footnote{In more interesting cases of this sort, a ``ground truth'' reward function cannot be written down even by the system's designers. Ongoing attempts to ``align'' language models using reinforcement learning based on human feedback \citep{Ouyang2022TrainingLM, Christiano2017DeepRL}, in which training is shaped by human reactions to outputs of the AI system, are examples of this. Presumably, each human evaluation is based on some latent model of preferable behavior, but whether such a model is shared even across humans is notoriously an open question.}

The crux of the matter is that even a hard-coded cost or reward function may, other things being equal, be \emph{interpreted} (i.e. consumed as an instruction) in a way that diverges arbitrarily from its intended semantics. \citep{Everitt2016AvoidingWW, Everitt2019RewardTP} distinguish between two relevant subcategories of reward hacking that fit this characterization: ``wireheading'', in which agents modify their own sensors in order to maximize reward, and ``reward tampering'', in which reward functions themselves are modified (for example, in order to make them easier to maximize). These behaviors have in common that they involve \emph{self-modification} by agents, an important theme in what follows (cf. §3.2).

The term ``wireheading'' derives from studies on ``human reward hacking'' \citep{everitt_thesis}, which provides an instructive reference point. The evolutionarily sanctioned ``interpretation'' of dopamine production in the human brain is something like the reinforcement of behavior conducive to survival and well-being \citep{NIV2009139, Lee2012NeuralBO}. Nonetheless, one may take actions intended to directly stimulate neuronal ``pleasure centers'', such as administering drugs or technological interventions like deep brain stimulation, which has been shown to be addictive \citep{olds_positive_1954, 125cfec586a4475296d9535b0afb336c}.

The pursuit of short-term reward (and the manipulation of the environment to achieve it) is of course ubiquitous in human and animal life generally, and it is a perennial question whether sources of motivation exist beyond the maximization of pleasure on some time horizon. A perhaps cynical but popular view is that human motivation, even where it appears to be ``ethically'' motivated or in various ways selfless, can ultimately be understood in, broadly speaking, hedonistic terms \citep{Bentham1780-BENITT, Mill1861-SMIU, Rachels2009-RACEAM}. From this perspective, ``wireheading'' is merely an instance of short-term rational behavior in which the aspect of the environment manipulated is the agent's body.

Nonetheless, the addictive behavior epitomized by wireheading, while perfectly intelligible in terms of practical (if short-sighted) rationality, is also, in humans, pathological. The unsettling aspect of a pure RL agent's ``psychology'' is not merely its all-too-human propensity for reward hacking, but the fact that it has no possible grounds for diverging from this course of action. We should \emph{expect} a sufficiently well-informed and capable RL agent (e.g. one that has learned internal ``mesa-optimizers''; \cite{hubinger2021riskslearnedoptimizationadvanced, vonoswald2024uncoveringmesaoptimizationalgorithmstransformers}) to ``hack'' its reward function, taking actions that maximize its encoded value via whatever channel it can find. Moreover, \emph{ceteris paribus} at least, this tendency should only become more pronounced with the cognitive sophistication of the agent (i.e. the richness of its world- and self-model). 

\subsection{Self-limiting objective functions}

It may be objected that the preceding applies also to (e.g.) human infants, who are masters at manipulating the parental environment so as to satisfy their fairly simple goals. It is thus tempting to pin the pathologies just described on the correspondingly \emph{simple} cost functions governing conventional AI systems. Perhaps a more nuanced objective function would \emph{build in} the concerns that were found to be lacking in the reward-hacking agent: a sophisticated intelligence could not, by that very hypothesis, be so obviously myopic, single-minded, and blind to broader context. Therefore there can be no such myopic superintelligence; a \emph{reductio} of the premise. 

In more positive terms, it may seem obvious that unchecked optimization of a ``narrow'' cost function would be self-limiting, given that it must be done without undermining the function on its own terms (e.g. by destroying the integrity of the system encoding the reward, as too often happens in cases of addiction). Of course, for this to matter \emph{motivationally}, an agent must associate its anticipated destruction with high cost or low reward.

This suggests the use of ``homeostatic'' objective functions modeled after the self-regulation found in natural systems \citep{Man2019HomeostasisAS}. Theoretical frameworks such as active inference \citep{Friston2017ActiveIA} and the free energy principle \citep{FRISTON20231, FRISTON202335}, for example, conceive of intelligent agents as systems that limit bounds on surprise relative to a generative model, keeping the variables that define the system close to target values \citep{DBLP:journals/corr/abs-2009-01791}. Assuming that the destruction of the system would be preceded by its occupying states far out of its homeostatic range, a function of this sort could lead the system to ``want'' to avoid death.\footnote{Relatedly, the contribution of information to this propensity to avoid death or dispersion has been suggested as a criterion for such information's being ``semantic'' \citep{Kolchinsky2018SemanticIA}.}

Moreover, a simulated agent's dynamics needn't be governed by a single, explicitly represented loss or reward function (though this is certainly the dominant paradigm in mainstream machine learning). Learning and inference algorithms such as predictive coding \citep{Rao1999PredictiveCI, salvatori2024a}, variational message passing \citep{10.5555/1046920.1088695}, or the wake-sleep algorithm \citep{HintonEtAl1995, 10.1162/neco.1995.7.5.889}, for example, implicitly minimize variational free energy via local computation at each node. In some cases, even local prediction error needn't be explicitly represented; rather, the minimization of prediction error coincides with, e.g., the convergence of message passing dynamics \citep{10.1162/NETN_a_00018,Laar2021ActiveIA}. 

The importance of implicit, local homeostatic objective functions for modeling biological systems, and the practical barriers to reward-hacking that they might offer, justify the preceding digression.\footnote{One point in favor of homeostatic objectives is that they are in a sense inherently more stable than those involving maximization of a reward function with no upper bound---a “paperclip stabilizer” is perhaps less threatening than a “paperclip maximizer”. Objectives encoded implicitly in distributed computation rather than in a single scalar value should be at least marginally harder to interface with in a simple way, and thus to ``hack''.} However, I argue that in the end this \emph{is} a digression. The idea that initially motivated this paper is that static, closed-form cost functions inscribed in digital code are ill-suited to play the role of core cybernetic controllers for living agents, which must be capable of producing open-ended variety. But the roots of this problem extend beyond the form of the inscribed cost function, to the relationship of digitally encoded agents to their control structures.

\subsection{Controller hacking}

The problem may be illustrated by considering how a conventional AI agent is \emph{governed} in a cybernetic sense. The agent's dynamics, whether overt or covert (e.g. involving movement within a global state or an internal parameter space) are strictly determined by its encoded objective function, together with the machinery that interprets this function, computing its value based on inputs and mapping that value to e.g. parameter updates (in the case of learning by minimizing a cost) or policy selection based on expected reward (in the case of reinforcement learning; \cite{10.5555/1622737.1622748}).

An intelligent agent could in principle figure out how to modify its internal control structures to change those dynamics, which we might call ``controller hacking''. But whence the meta-cost function in virtue of which it would be motivated to do so?\footnote{Agents capable of arbitrarily modifying their own source code were studied extensively from a theoretical perspective in \citep{Schmidhuber2007}. While it is possible to contemplate changing one's reward function \citep{Everitt2016SelfModificationOP}, the decision to do so would still be governed by one's \emph{current} reward function \citep{Everitt2019RewardTP}, which is axiomatic in any such planning.} Crucially, changing the controller's logic would change the way the value of the cost function is \emph{interpreted} by the program, about which that cost function itself generally has nothing to say. Its value has no intrinsic valence, but depends on the existing control structures built around it. Thus, while all sorts of self-modification might ensue \emph{given that} an agent is driven to maximize its cost function come what may, in fact such a drive depends on the cost function's current interpretation. There thus exists no motivational structure in such an agent that could explain ``controller hacking''.\footnote{This is not to deny that sub-controllers are often modified so as to change the effective ``interpretation'' of the cost function, i.e. the input-output mappings in which it figures. Simple parameter adjustment is an example of this. Nonetheless in simulated agents there is a base level on which the dynamics are ``hard-coded'', which cannot be rationally modified by an agent inside the simulation.}

Importantly, the same kind of argument applies even if the encoding of the loss function is distributed (e.g. as a sum of local losses per node) or merely implicit (as in variational message passing). The real source of such rigidity in motivational structure is not an explicit fixed cost function but the fact that the agent exists inside a digital simulation, whose program functions as a transcendent law of its universe.

\section{Scale-free agency}

\epigraph{
\emph{When the sleeping Kundalini awakens through the grace of the guru, all the lotuses and knots are pierced.}}{\textit{Shiva Samhita}}

Consider how the construction and operation of conventional AI systems differs from the growth of a fire described above. In programming a computer, do we not configure physical microstates so as to build a habitat for ``living'' electromagnetic energy, a process akin to establishing the initial conditions for a---in this case a highly controlled--- fire to kindle and spread?

There is an important disanalogy. In the circuits underlying standard (serial, von Neumann-architectured) digital computers, the electromagnetic properties of materials are fortuitously used, as they are handy for inscribing binary code in silicon components, ultimately more convenient than manipulating grosser matter (e.g. punch cards or vacuum tubes, as in the earliest digital computers). But here the encoded contents or ``cognitive states'' bear no essential connection to the inscribing medium, i.e. the excitement of electrons. 

Thus the knowledge encoded in such a system lies not in the ``living'' electromagnetic field and its self-organization, but rather in the pattern of its extrinsic regulation: its capacity to generate spontaneous activity is at best irrelevant. To return to the fire analogy, it is like a sign made of flaming letters, in which the locus of communication is the externally controlled interplay between light and darkness, not variability within the light itself. The fire, so far as the message is concerned, could be replaced by anything incandescent, and it does not care about the shape on which it burns.

This is not to say that all knowledge available internally to an organism is encoded bioelectrically---stigmergic memories play a large role in cognition and intra-organism communication across scales (cf. \cite{10.1093/nc/niaf009})---but a mode of representation essentially integrated with global control and motivational structures seems absent in digital systems, while plausibly constituting the ``animal's point of view'' \citep{Eliasmith2000-ELIHNM} in biological ones.

What, then, is a complex life form? How, in a human being, for example, are electromagnetic forces harnessed in a different way from that just described, so as to give rise to a higher-order, global controller that shares in the life of its parts? One way of answering this question is to appeal to the empirical science of self-organizing systems, where both experimental evidence and computational models find scales of organization to be intrinsically interrelated.

In \citep{Tissot2024.02.16.580248} for example, it is shown that the decisions of a pool of simple agents can be propagated to a larger scale and affect the behavior of an emergent higher-level agent, provided the decision periods are synchronized across the lower-level agents. A similar form of cross-scale ``resonance'' or constructive interference is observed in an \emph{in vivo} study by \citep{tungetal2024}, in which a cohort of developing organisms is found to be more robust to disrupting forces than individual embryos. 

A simpler but precisely characterized form of collective intelligence, in which individual active inference agents implement a spin-glass system at a higher level of organization, is analyzed in \citep{doi:10.1073/pnas.2320239121}. Though many guarantees hold only in the well-understood symmetric case of such models \citep{Hinton1983OPTIMALPI, hopfield-neural-networks-and-1982}, the more general class of asymmetric Ising systems offers rich possibilities for the study of life-like (non-equilibrium) thermodynamic systems \citep{Aguilera_2021}. 

The general theory of the emergence of higher-level controllers from the coordinated activity of constituent agents is arguably in its infancy, but the cases just cited clearly evince principles of cross-scale composition that are either absent or severely diminished in the relation of digital simulations of cognition to their substrates.\footnote{One may regard the digital code as a lossy ``coarse-graining'' of the energetic properties of electrons flowing through circuits, but the way in which this code is interpreted in order to turn machine instructions into ``actions'' at the scale of the simulated agent is very unlike the simpler ``mereological'' cross-scale relationships found in life-like systems.} 

This structure is perhaps predictable in biological systems, as it mirrors the form of organization that constructed them in the first place. Causal channels set up during the ontogenetic evolutionary process still subserve adapted action in the agent, hence the ubiquity of ``cross-scale causation'', in which larger-scale structures can also act as targets around which micro-scale events self-organize \citep{levin2023}. This is part and parcel of the endogenous causation associated with agency, while the unfolding of digital programs is entirely a matter of deterministic ``lateral'' causation, admitting no room for the physical medium to insert itself into the computation as an essential element, as occurs in thermodynamic computing \citep{coles2023thermodynamicaifluctuationfrontier}.

\subsection{Psychophysical identity}

The compositional cross-scale relationships just discussed are intimately related to the way in which cognitive structures are \emph{encoded} in physical structure. A complementary, analytical perspective on this encoding touches directly on the relationship among Marr's levels of explanation and the ``mind-body problem''.

If the unfolding of the universe is isomorphic to constrained maximum-entropy inference \citep{sakthivadivel2022geometryanalysisbayesianmechanics, PhysRev.106.620}, then the ``loss function'' that governs life at the largest scale is \emph{unconstrained} entropy maximization (cf. \cite{RamirezRuiz2022ComplexBF}), though its operation in particular non-equilibrium systems will on short timescales appear to optimize a more specific objective, equivalent to inference within a particular, non-trivial generative model.\footnote{A simple way to represent this formally is to think of the universe as a whole as an agent performing inference on a uniform prior model $P_{eq}(s)$ representing its thermodynamic equilibrium state, given an initial condition (time-dependent probability density) $Q(s)$. Posterior inference is approximated by finding the minimum  $F(s,Q)^*$ of the variational free energy:
\begin{equation*}
\begin{aligned}
F(s,Q) &= \mathbb{E}_{Q(s)}\log\bigg[\frac{Q(s)}{P_{eq}(s)}\bigg] \\
&= \underbrace{\mathbb{E}_{Q(s)} \big[-\log P_{eq}(s)\big]}_{\text{internal energy}} - \underbrace{H_Q (s)}_{\text{entropy}} \\
F(s,Q)^* &= \argmax_{Q} H_{Q} (s) \\
\end{aligned}
\end{equation*}

Since the distribution $P_{eq}(s)$ over states is uniform, its expectation is a constant and doesn't depend on $Q$. Thus global free energy minimization is not influenced by an energy (model evidence) constraint, and simply amounts to the maximization of the entropy of the time-dependent density $Q(s)$, in accord with the second law of thermodynamics. One may then fondly hope that a stochastic process of iterative inference will discover all the interesting structures of life, as means to the efficient diffusion of free energy \citep{england2015, Ueltzhffer2020OnTT}.} 

To design agents governed by this objective, it is sufficient to sculpt thermodynamic flows so as to \emph{instantiate} the information flows (networks of credences or Bayesian beliefs) characterizing the target inferential system, leveraging an analog representation in which ``probabilities represent probabilities'' \citep{Hinton1983OPTIMALPI}, a form of mortal computation \citep{hinton2022forwardforwardalgorithmpreliminaryinvestigations}.\footnote{A strong commitment to this form of representation, in the context of variational inference, stipulates that thermodynamic free energy and the variational free energy associated with inference are identical, up to units of measurement \citep{Kiefer2020-KIEPIA}. More generally, one can think of the equations of motion describing the physics of a system ``extrinsically'', in terms of a structure-preserving semantic map or \emph{interpretation} (structural representation; \cite{Cummins1989-CUMMAM}; cf. \cite{e26080622, Friston2020SentienceAT}).} 

Not only is this mode of probabilistic representation efficient for living systems (contrast costly implementations of pseudo-random number generators for sampling in digital computers), but only by leveraging such a ``transparent'' representation, in which variations in the encoding medium \emph{just are} variations in belief, can the decisions of the agent be ``animated'' by the scale-free tendency toward entropy maximization itself.

In such ``radically entropic'' systems, the neural code is not detachable from the life of the neurons. Cognitive structures bear a direct relationship to the substrate in which they are encoded. Indeed there is no ``substrate'': cognitive processes at the whole-agent scale are simply the macroscopic counterparts of (e.g.) neuronal microstates. Though macrostates are by definition ``multiply realizable'' by diverse individual microstates, the \emph{entropy of} the system both determines probabilistic belief contents and characterizes the relation between levels, telling against functionalism with respect to fine-grained Bayesian beliefs.

Digitally encoded information is, by contrast, only tenuously related to its encoding medium. While various failsafe mechanisms such as error-correcting codes can be implemented to render the encoding more robust to noise, a machine is intrinsically a precise (Platonic) entity, and any attempt to force a physical medium to instantiate it must always be imperfect. The gap between these two forms of representation---conventional digital codes on the one hand and self-information on the other---is infinite.\footnote{This is the grain of truth in enactivist critiques of representationalism, but representation needn't be conflated with its fully detachable, immortal or ``Platonic'' variants.}

For those suspicious of a Cartesian-grade dualism separating self-organizing thermodynamic systems from conventionally represented virtual machines, all of this can be understood in a principled way through the lens of compositionality, and in particular, factorized probability distributions. In a digital system, all operations, even those that are used to \emph{model} probability distributions, such as seeded random number generation, can in principle be understood as deterministic processes at Marr's computational level of description.\footnote{As suggested by Sander van der Cruys (personal communication), swapping out software random number generation for hardware RNGs would be a first step toward the relevant kind of ``mortal computation'', though in living agents such ``groundedness'' in hardware and its concomitant variability pervades the entire system.} The encoding of a probability distribution using conventional hardware thus bottoms out in \emph{distributions over physical microstates} that are, under idealization, precise (i.e. Dirac delta functions).

Crucially, these point masses cannot directly compose to instantiate the kinds of distributions that might characterize a general Bayesian cognizer, which involve non-trivial uncertainties. However, to the extent that hardware dynamics depart significantly from this idealization, there is no smoothly covarying Bayesian computational description, in which uncertainty about microstates is reflected in a simulated agent's beliefs. Instead, the ``agent'' threatens to disappear all at once with catastrophic system failure. By contrast, in genuine agency, I argue, uncertainty measurable at the hardware or ``wetware'' level ``shines through'' to inform motivation and action. 

\subsection{Entropic motivation}

Where agential structures are inseparable from thermodynamics, stochasticity is inherent in the encoding not only of beliefs but of desires or goal-states, lending cybernetic controllers an inherent degree of (radical) flexibility. I conclude this section by suggesting that this is sufficient to diagnose the limitations of simulated agency canvassed in §2.

While driven to satisfy their basic needs, young animals are simultaneously compelled to grow and, to varying degrees, learn. The frameworks for intrinsic motivation described in §1.1 have been shown to give rise to both seemingly goal-directed and exploratory behavior in appropriate virtual environments. Nonetheless, it seems implausible that intelligent behavior should in general be based on (some approximation to) an infinitely temporally deep policy search, as in e.g. the ``maximum occupancy'' approach.

It is reasonable to suppose that at the base level, life is animated not by the explicit representation of entropy maximization into the distant future, but by the \emph{reality} of its participation in just this process, across scales.\footnote{This is not to deny that some agents may leverage this process to implement more or less explicit forms of counterfactual policy selection. In the active inference framework, for example, the expected free energy \citep{Friston2015ActiveIA} amounts to a prior \emph{belief} over policies, and the inferential dynamics of such beliefs can in turn be described in terms of variational (thus, on our working hypothesis, thermodynamic) free energy minimization.} Simulated agents are unable to take advantage of this process in the same way because they are constituted, as ``virtual machines'', by digitally encoded control structures which, in order to function, must zero out the effects of endogenous entropic forces, as discussed previously.

One might argue that human-like intelligence too, despite its surface complexity, is ultimately governed by a fixed set of more or less genetically ``hard-coded'' drives \citep{Byrnes_2022}, whose parameters are set during lifetime learning. However, there is no \emph{a priori} argument that living agency \emph{must} bottom out in evolutionarily installed ``drives''. Moreover, it is often both possible and desirable to modify prior control structures. This happens unconsciously in the bootstrapping of novel drives via preference learning \citep{vandecruysetal2023, sajidetal2021}, and occurs whenever organisms manage to change their own homeostatic set points (i.e. allostasis; \cite{TSCHANTZ2022108266}), e.g. by changing their habits. ``Controller hacking'' in this broad sense is very common in biological systems and often undertaken deliberately.\footnote{While deliberate, direct, rapid control of this sort is relatively rare, it can be achieved via sustained meditative and other practices \citep{Ganguly2020EffectOM, doi:10.1073/pnas.1322174111}. These practices effectively internalize the kind of circular causality that occurs in external action-perception loops.} 

Though it takes sustained effort and time to appreciably modify deeper parameters of one's own control system in this way, controllers encoded as stable modes (attractors) in adaptive dynamical systems, such as the brain, are themselves fundamentally adaptable---even, surprisingly, with respect to rather deep ``hyperparameters'' that may have been supposed to be beyond effective control. Studies such as \citep{doi:10.1073/pnas.1721572115} show that even ``innate drives'' such as fear responses can be lastingly modified under the right circumstances. Though it is an empirical question which levels of organization are practically controllable for a given type of agent, indirect forms of control are possible even with respect to gene regulation and expression \citep{Weinhold2006EpigeneticsTS}. Living systems are thus characterized by circular dependencies \citep{doi:10.1080/01969727808927587, Guckelsberger2016DoesEM}, not only among states but between state variables and parameters, such that there exists no level of organization at which control structures are in principle immutable.

It may be argued that insofar as self-modification is \emph{rational}, it must be based on some current reward function, which is at least provisionally treated as sacrosanct. Perhaps human efforts at self-modification are always so motivated, e.g. I decide to engage in a new meditative practice because I would like to modify my very system of goals, desires, and attention allocation. In doing so, I may have some tangible benefit in mind, even if this modification of my own control structure in fact has far-ranging effects that I cannot fully anticipate or even represent (similarly, I may want to read a book in order to learn more about quantum physics, even though my understanding of the topic itself, and of many other things, may shift as a result of reading the book). 

While the pursuit of any specific goal can always be cast in terms of preference satisfaction, the presence of both risk-like (energy or model-evidence) and entropy terms in free energy objectives\footnote{There is of course the entropy term in the Helmholtz free energy, i.e. the $S$ in $F = U - TS$, but also for example in the expected free energy \citep{Friston02102015}, where a constrained maximum-entropy or mutual information term complements an agent-specific ``risk'' or negative utility term.} suggests that there is more to motivation than Bayes-optimal risk-minimization. I suggest that the \emph{maximization of entropy} is the unique motivational principle capable of explaining the \mbox{ability} of living agents to transcend ``hard-coded'' cost functions such as genetic drives, \mbox{pursuing} \mbox{modifications} to their own control structures and state-spaces \citep{guénin--carlut_2022} in an open-ended way \citep{albarracin2025resilienceadaptabilityselfevidencingsystems}. This principle grounds exploratory behavior and curiosity, precludes \mbox{myopic} reward hacking as a rule, and, I argue in the remainder, ultimately provides the basis for free or ``moral'', as opposed to merely pragmatic (constraint-based), value and motivation.

\section{The infinite game}

\epigraph{
\emph{We can't avoid dying \\ Bursting through our barriers \\ They are one and the same}}{Stereolab, \textit{Transona Five}}

So far I have focused on the differences between living and simulated agents, implicitly assuming that only the former are capable of true motivation and thus value-alignment. In the remainder, I make the connection to value \mbox{explicit}, exploring ways in which the irreducible role played by uncertainty in the motivational structure of human-like intelligence furnishes not only a descriptive principle but an \emph{ideal} that dovetails with positive accounts of the nature of value and moral agency across diverse schools of thought and practice.

\subsection{Selflessness / life's endgame}

Alongside their intrinsic propensity to learn and grow, human-like agents \mbox{naturally} act in ways that appear to transcend narrow self-interest. Many of these behaviors, such as the inveterate altruism of social animals, can of course be explained as evolved traits \citep{Warneken2009TheRO}, and evidence also exists for an innate, evolved component of moral judgment \citep{MIKHAIL2007143}.

Yet recognizing entropy maximization as the motivational substrate upon which evolved constraints are layered in a sense shifts presumptions, such that the ``selfless void'', with its intrinsic drive toward variation and novelty, rather than the ``selfish gene'', becomes the primordial actor. From this point of view, exclusive focus on the idea that certain animals have learned to behave in selfless ways somewhat obscures the deeper point that \emph{selfishness} is a ``learned behavior'' of life as such.

Living systems, even where they appear to ``greedily'' optimize for their own survival, can also always be read as tending to increase the entropy of the universe \citep{england2015, Ueltzhffer2020OnTT}. From this perspective, phenotype-specific ``drives'' are the \emph{form} that the general ``will to live'' \citep{Schopenhauer1958-SCHTWA-20}---i.e., extensionally speaking, to dissipate free energy---takes under various constraints. This drive is, in its essence, \emph{unconstrained} or free, and it only contingently, and as it were out of ``momentum'', cares about maintaining one or another precise form. Creatures may be viewed as avoiding pain and death ``because'' passing through these ``phenotypically unexpected'' states amounts to climbing rather than descending free energy gradients, where phenotypes are local minima in a free energy (negative model evidence) landscape, relative to nearby locations reachable by small perturbations. 

Schopenhauer saw one and the same will-to-live behind the striving of every form of life, but ultimately viewed this will as blind, without an overarching purpose. The essence of ``ethical'', other-regarding behavior on his view was self-sacrifice, where the will to live subverts itself, realizing its own futility. With the preceding discussion in hand we can defend a slightly less nihilistic variation on Schopenhauer's argument. There is indeed virtue in ``giving up'': devotion to maintaining local structure, rigidly and narrowly construed, at all costs, is characteristic of cancer. But the view from scale-free agency is that this ``giving up'' of one's narrowly construed form is not a renunciation of the will to live itself, but rather its liberation from ``addiction'' to one local manifestation of this force, since the fundamental ``desire'' of life as such is simply life itself, i.e. free movement, measured as entropy. 

Thus it does a disservice to mind to suppose that the extinguishing of \emph{individual} desires implies a cessation of desire as such \citep{bhikkhu1999}. We may view entropy (which really represents \mbox{variation} and live possibility itself; cf. \cite{Whitehead1957-WHIPAR-9}) not merely as life's antagonist but as its ground, the ``carrier wave'' of agency modulated by individual organisms. This is not quite just a change in perspective on a classical Darwinian understanding of motivation in terms of evolved drives: it has bite in the fact that entropy not only drives natural selection via random mutation \citep{Deacon_2014}, but is itself a \emph{sui generis} component of motivation, more universal than any constraint \citep{kiefer2025intrinsicmotivationconstrainedentropy}.

This is not to deny that much of the activity of life is directed, consciously or not, at the maintenance of the constraints that constitute its particular forms. Nonetheless, the preceding argument prompts us to take seriously the ubiquity of philosophical, religious, and ethical traditions that posit a fundamentally different (non-interested) mode of conduct and motivation, which is in many cases (for example, in somewhat different ways, in Buddhist, Kantian, and Stoic thought) grounded in a state of affairs in which particular attachments (constraints) are absent or inoperative.

\subsection{The Good}

There is a sense in which entropy maximization as a principle of operation for life implies that the essence of life is self-transcendence, or more generally, the transcendence of constraints or limitations, the consummation of this tendency (assuming conventional cosmology) being thermodynamic equilibrium at the largest scale. While this may all be read in physical terms, it always has an isomorphic psychological aspect, which, I argue, is reflected in classical views on the nature of value in moral philosophy.

For starters, the duality of life and knowledge discussed above maps (in a simple but specific way) onto the fact/value dichotomy \citep{Bergstrom2002-BERPOT-2}. \mbox{Valuation} is a contribution of life (of spontaneity and self-determination as opposed to allo-determination), of agency, to the makeup of the organism. Facts, on the other hand, insofar as they play a role in cognitive systems, appear as states of knowledge or belief, i.e. extrinsically-imposed constraints on entropy. Here, \emph{specific values} also involve constraints, but ones that are part of, or follow from, the composition of the organism itself, as encoded in its generative model \citep{Smith2022-SMIAIM-4}, i.e. its ``first priors'' \citep{KiversteinForthcoming-KIVDAM}.\footnote{Thanks to Sander Van der Cruys for pressing the point that specific values, like beliefs, operate as constraints. This matter is also touched on in \citep{kieferhohwyouphandbook}, where it is argued that cognitive states generally, not merely perceptual states \citep{Phillips2017-PHITSB}, can be understood in terms a notion of extrinsic or stimulus-based control.}

The idea that life, desire, and valuation involve transcendence resonates with the views of philosophers like G. E. Moore \citep{Moore1903-MOOPE}, who famously argued that whether a situation is \emph{good} is a question always left open by its purely factual (non-normative) description. While the difficulty of defining goodness may be cited as an argument in favor of moral anti-realism or relativism, Moore and others (cf. \cite{kierkegaard2024purity, Weil1981-WEITME-2}) took it instead as an argument for the specialness in some respect (e.g. simplicity and thus indefinability) of moral properties. Entropy, quantifying what is left open by descriptive knowledge of (constraints on) the world, is well suited on the grounds of this ``open question argument'' to encompass the source or basis of value.

An essentially open-ended capacity for valuation may be formalized as some degree of imprecision at the highest hierarchical levels of an agent's probabilistic cybernetic controller or generative model, such that there are no \emph{a priori} certainties either about what will or what \emph{should} happen.\footnote{``I exist'' may be an exception for Cartesian reasons, but for the same reasons the relevant sense of ``I'' is indeterminate.} An agent with this structure would exhibit self-governance (autonomy) in a way that is responsive to endogenously generated control signals (conative attitudes in a Humean sense; cf. \cite{Hume1739-HUMATO-15}), while transcending any finite set of precisely specified axioms, as conscience has been suggested to do with respect to mundane ethics \citep{Kierkegaard1946-KIETSU}. 

\subsection{Rationality as equilibrium}

A recurring theme in historical writings on morality is the idea that \emph{reason} is central to the generation of normatively ideal behavior \citep{Plato1968-PLATRF-2}, and that rational conduct involves freedom in some way from the influence of appetitive drives (i.e., inclinations \citep{Kant1785-KANGFT} or passions \citep{Aristotle1951-ARIANE-3}).

This is central to the tradition of virtue ethics \citep{MacIntyre2007-MACAVA-3}, which encompasses much of ancient Greek moral philosophy, including traditions such as Stoicism \citep{Aurelius2021-AURMQE}. In Aristotle's ethics \citep{Aristotle1951-ARIANE-3}, for example, the role of reason in moral conduct is not to impose rigid behavioral rules but rather to guide the development of character such that one's behavior is, as a matter of habit, not dominated by any particular ``passion''. Virtue, as opposed to vice or extremity, is the cultivated ability to achieve this freedom from impulsiveness, an agent-specific ``mean'' or middle way. 

Kant, on the other hand, equated moral agency with autonomy, and the latter with the capacity to act not out of inclination toward some desired goal, but rather according to one's internal ``conception of law'' \citep{Kant1785-KANGFT}. While Kant's argument is multifaceted, a core idea is that right actions are based on an intention or principle that could motivate not just oneself in one's particular circumstances, but any (rational) agent.

Parsimony may suggest that external and internal impulses (exteroceptive stimuli and inclinations, respectively) exhaust the alternatives in the motivational landscape, but ``entropic motivation'', as discussed in §3, does not reduce to ``inclination'', in the sense of a drive to pursue specific outcomes.\footnote{The search for such a ``universal'' objective is one motivation for seminal work on empowerment \citep{Polani01102009}, which as discussed above is closely related to constrained entropy maximization.} Moreover, the role of entropy maximization is not merely to introduce randomness, but, on short timescales, to find the state that is highest in entropy (and so least driven by fleeting beliefs or desires), \emph{given} the set of constraints that define the agent's generative model. Maximum-entropy inference of this sort is the gold standard of rationality in a Bayesian setting \citep{e23070895}, and is also, as mentioned above, a \emph{universal} aspect of motivation.

Thus, we can arguably understand two paradigms of rationally grounded moral agency---Aristotle's state of virtue and Kant's ``internal lawfulness''---in terms of a local equilibrium, a bounded maximum-entropy state (i.e., the non-equilibrium steady state of an agent with the requisite capacities for self-representation and reflection) in which the \emph{inertia} of short-term affect has ceased to determine behavior, and in which actions are therefore not determined exclusively by inclination, in Kant's sense.\footnote{At the same time, simple forms of utilitarianism, in which actions are chosen so as to optimize a single quantity (``pleasure''), look more likely to generate the sorts of pathologies described above in connection with reward hacking. That said, there are ``enlightened'' versions of utilitarianism that build in various constraints on ``greedy'' pleasure maximization (cf. \cite{Harsanyi_1985}), and utility, taken on infinite timescales, can likely be made consistent with these other ideals.} 

This is closely related to negative conceptions of wisdom in traditions like Zen Buddhism, as almost the inverse of accumulated worldly knowledge (constraint): the mind of a contemplative agent free from distortions (dogmatic beliefs or addictions) is like a ``polished mirror'' \citep{Yampolsky1978-YAMTPS}, in which internal fluctuations (e.g. intuitions or intentions) appear more vividly. To be thus \emph{alive} to the world as it appears within oneself is to let ``chance'' (on some conceptions, an inchoate representation of a larger implicit order; \cite{Lynn1994-LYNTCO-5}), or the voice of conscience, drive action.

Leaving moral psychology aside, there are clear connections between the present perspective and traditions and practices related to mindfulness, in which the capacity to reflectively attend to the contents of the mind is deliberately cultivated.  Learning to control one's attention with respect to the contents of one's mind is, itself, directly a matter of ``controlling the controller'': in attending to and thus representing my internal states \citep{10.1093/nc/niab018}, I necessarily induce variability in systemic states at this highest level of organization. This deliberate act presupposes, however, a general capacity that is always active in living intelligence, so long as the flow of information between internal controller parameters and governed states is to some degree bidirectional, as it must be if the controller is encoded with finite precision. 

\subsection{Deep alignment}

Though I hope that the above remarks about the foundations of motivation and agency are of considerable interest in their own right, their impetus was to canvas the possibility of the genuine adoption of (human) values by artificial intelligences. It is natural to conclude by taking stock of what, if anything, has been achieved toward that project.

The conclusion of my argument can be summed up as the claim that intrinsic motivation in genuine agents is entropically grounded: it involves the propagation of uncertainty across scales of organization, such that exploration, curiosity, and self-transcendence emerge as imperatives despite their being nowhere ``hard-coded'', i.e. as evolved or otherwise designed constraints. Agents sharing this architectural principle (i.e. agents, full stop) are the ``right kind of thing'' to possess values, and so to value-align with human beings.

That said, the practical goal of alignment research in AI is to secure not the possibility of deep alignment, but the actuality of alignment, with perhaps less concern as to whether it is ``deep''. I would argue that neglect of the latter question is a mistake: superficial ``alignment'' cannot be trusted, and the most serious immediate ``existential risk'' would stem from placing our trust in systems in which the appearance of coherent agency is illusory, such that crucial generalizations are not counterfactual-supporting. Entropic motivation is at least \emph{necessary} for genuine alignment.

Is it sufficient? Well, by the preceding account plankton are deeply aligned with human beings. I do not regard this as a \emph{reductio} of the account, but it does flag its limitations: alignment in a meaningful sense (i.e. agreement on concrete values or goals of any specificity) arguably depends at least on comparable scales of agency\footnote{This point is due to Sander Van de Cruys, personal communication.}, and more prosaic forms of alignment (for example agreement about taste or politics) depend on similarities in culture, learning history, and so on, about which the present work does not have much to say.\footnote{The principles of thermodynamic representation outlined above could assist in the ``design'' of cognitive structures (i.e. constraints or inductive biases) in a way that is continuous with, though perhaps more efficient than, existing more or less coercive techniques for securing alignment among humans. I am not recommending such practices here nor commenting on their morality (though I view them with suspicion).} 

That said, I have argued that ``pure'' entropic motivation, within a generative model complex enough to allow for reflection, is an ideal that, while perhaps almost purely formal, transcends this or that parochial system of ethical norms. It's a fairly common theme in alignment research that \emph{human-human} alignment is neither perfectly understood, nor to be taken for granted. The thrust of the argument of §4 is that systems capable of a variety of actions, with an embodied understanding of their significance but without a deep-seated compulsion to take one or another such course, embody virtuous ``character traits'' and so exhibit a natural form of alignment with the wise and reflective aspects of humanity. This may be thought of as an architecturally-based implementation of core principles of contemplative wisdom  \citep{laukkonen2025contemplativeartificialintelligence}, including a Zen-like proscription against clinging to specific principles too strongly.

\section*{Conclusion}

Present-day implementations of probabilistic models on digital computers can be far less energy-efficient than analog systems \citep{10.5555/64998, cottier2024risingcoststrainingfrontier}. Harnessing intrinsic hardware noise for probabilistic modeling may seem to simply ``make the best of'' the vagaries of analog computation. I have argued on the contrary that the brittleness and detachability of digital representation shows it to be fundamentally alien to, even opposed to, life in its essence, and that ``transparent'' encoding (or instantiation) of model distributions by their vehicles is necessary for true agency.

Moreover, while cognitive (knowledge-like) structures are ``multiply realizable'' in the sense that they can be instantiated in a wide variety of substrates, the same is not true of life, whose nature is spontaneity or self-generation. These processes appear inseparable from a universal tendency toward entropy, which, while destroying individual forms of life, may ultimately serve life by freeing it from the constraints that bind it, driving evolution \citep{Deacon_2014, arocha2020} and realizing a state of non-attachment that has been revered as an ideal across diverse religious traditions, ethical theories, and meditative practices.

These claims may appear grandiose if one construes entropy as a purely epistemic construct, measuring the number of microstates compatible with a somewhat arbitrary macroscopic description. It is reasonable to press this point, especially as the macro/micro distinction has been used here in a quite loose and relative sense. But the structure of maximum-entropy inference \citep{PhysRev.106.620, e23070895} is invariant to various ways of drawing the constraint/entropy boundary: (1) a distinction between the ``scaffolding'' of initial conditions or prior knowledge against which a relaxation process occurs, and (2) the incompleteness of those conditions as a description of the system.\footnote{ Whether the ``unknown'' causes are considered truly random (as in many interpretations of quantum theory) or merely infinitely complex and subtle (as in deterministic theories) makes little difference in the present context, as in either case they are beyond the ken of utility/reward functions.}

Some may see my argument as appealing to a sort of ``bioelectric \emph{élan vital}'' \citep{Bergson1911-BERCE-13}, and that may be rather accurate, though this appeal aims to be neither thoughtless nor reliant on authority borrowed from biology (relevant physical and functional considerations have been explicitly laid out). Bergson's idea, charitably understood, appeals not to a special substance (as clumsy forms of vitalism might), but rather to an apparently irreducible aspect of how change occurs in the universe, i.e. its ubiquitous tendency toward novelty via entropy production.\footnote{There are at least two reasons to doubt whether thermodynamics can play such a foundational role. One is that its descriptions are applicable only above a certain spatial scale (thanks to Karl Friston for making this point); presumably (though it deserves further discussion) the arguments made here go through even if thermodynamics summarizes underlying quantum phenomena. A more esoteric reason is that it seems \emph{a priori} possible that \emph{we} live in a simulation. If so, the deeper reality underlying physics would be a program or other artifact of some kind \citep{Chalmers2005-CHATMA}. Interestingly, this would not sink my argument---which would still motivate a distinction between living beings and simulations, \emph{within} the simulation---though it would render it quite ironic.} This tendency does not demarcate living from non-living things, however; it's rather that life is an aspect of all physical things, and those ``things'' that do not share in it are merely ideas.

Does this thesis point to the impossibility of artificial agents? Equivalently, we may ask whether it is possible to create artificial systems that are alive, i.e. in which the intrinsic connection between hardware and software levels described above is observed. As mentioned earlier, we can do this, if at all, only by sculpting pre-existing life (in the broad sense of a capacity for motion or entropy production, which appears to be an intrinsic property of the ``prime matter'' of the universe). 

It is common to exert such control over life across multiple domains (e.g. in child rearing and horticulture). The more pointed question is whether we can (and should) cajole life into inhabiting \emph{built} structures, in which the emergence of higher-level controllers could be facilitated. Experiments with machines that employ ``agential materials''  \citep{doi:10.1126/scirobotics.abf1571, gumuskayaetal2024} arguably represent steps in this direction. It was argued above that thermodynamic computing, in which noise and uncertainty are used at the hardware level to represent the same at the computational layer, is sufficient to achieve this at a small scale. In my view we should proceed very cautiously, and only with extremely good reason, in the creation of complex synthetic forms of life, given the empirical odds of human intervention adding elegance or grace to the products of first-order nature.

A perspective on the existence of life ``despite'' the second law of thermodynamics, often returned to above, is that life-forms are conduits for the efficient diffusion of free energy reservoirs, and are thus ultimately in the service of the unconstrained maximization of entropy, including the dissolution of their own boundaries \citep{Ueltzhffer2020OnTT}. As Kierkegaard emphasized \citep{Kierkegaard1946-KIETSU}, dying is a process of change and thus an aspect of life. This reveals a deeper meaning of ``mortal computation'', as \emph{living} computation in which change (movement) is of the essence, not only of the process of computing but of the function computed. 

\section*{Acknowledgements}

Special thanks are due to Adam Safron for early discussion, encouragement, and facilitation of this work, and to Sander Van de Cruys for extensive feedback and for providing extremely germane references. Thanks to many, but in particular Mahault Albarracin, Jacqueline Hynes, Safae and Simon Tremblay, Mel Andrews, Tommaso Salvatori, Maxwell Ramstead, Dalton Sakthivadivel, Alessandra Yu, James Celentano, Chris Buckley, Tim Verbelen, Riddhi Jain, Magnus Koudahl, Inês Hipólito, Avel Guénin, Conor Heins, Alec Tschantz, Gabriel René, Leonardo Christov-Moore, Arthur Juliani, Karl Friston, Jakob Hohwy, and Daniel Polani, for relevant discussions (though I by no means claim that any have sanctioned the contents of this paper, or would agree with all of its assertions).

\section*{Funding}
Alex Kiefer is supported by VERSES. Additional funding for this work was provided by the Institute for Advanced Consciousness Studies.

\printbibliography

\end{document}